\documentclass[runningheads]{llncs}
\usepackage{multirow}
\usepackage{graphicx}
\usepackage{bibnames}
\usepackage{booktabs}
\usepackage{verbatim}
\newcommand\correspondingauthor{\thanks{Corresponding author: thindv@uit.edu.vn}}

\usepackage[colorlinks=true,linkcolor=blue,citecolor=blue,urlcolor=blue]{hyperref}
\begin{document}
\title{Software Mention Recognition with a Three-Stage Framework Based on BERTology Models at SOMD 2024}
\titlerunning{Three-Stage Framework Based on BERTology Models at SOMD 2024}
% If the paper title is too long for the running head, you can set
% an abbreviated paper title here
%
\author{Nguyen Thi Thuy\inst{1,2} \and Nguyen Viet Anh\inst{1,2} \and
Dang Van Thin\inst{1,2}\correspondingauthor \and Ngan Luu-Thuy Nguyen\inst{1,2}}
\authorrunning{Thuy et al.}
% First names are abbreviated in the running head.
% If there are more than two authors, 'et al.' is used.
%
\institute{University of Information Technology - VNUHCM \\
\and Vietnam National University, Ho Chi Minh City, Vietnam \\
\email{21521514@gm.uit.edu.vn}, \email{19521204@gm.uit.edu.vn}, \\ \email{\{thindv,ngannlt\}@uit.edu.vn}}
\maketitle              % typeset the header of the contribution
\begin{abstract}
This paper describes our systems for the sub-task I in the Software Mention Detection in Scholarly Publications shared-task. We propose three approaches leveraging different pre-trained language models (BERT, SciBERT, and XLM-R) to tackle this challenge. Our best-performing system addresses the named entity recognition (NER) problem through a three-stage framework. (1) Entity Sentence Classification - classifies sentences containing potential software mentions; (2) Entity Extraction - detects mentions within classified sentences; (3) Entity Type Classification - categorizes detected mentions into specific software types. Experiments on the official dataset demonstrate that our three-stage framework achieves competitive performance, surpassing both other participating teams and our alternative approaches. As a result, our framework based on the XLM-R-based model achieves a weighted F1-score of 67.80\%, delivering our team the 3rd rank in Sub-task I for the Software Mention Recognition task. We release our source code at this repository\footnote{https://github.com/thuynguyen2003/NER-Three-Stage-Framework-for-Software-Mention-Recognition}.


\end{abstract}

\section{Introduction}

% gioi Thieu ner task
Named Entity Recognition (NER) is an important task in NLP that involves identifying and classifying named entities in text. That will transform them into structured data, making it easier to categorize and perform search processing or carry out other NLP tasks \cite{dash2024clinical} on that data such as text classification, sentiment analysis, and contextual analysis, ... particularly in the domain of Biomedical Named Entity Recognition (Bio-NER), which is challenged by a range of entities like genes, proteins, medications, and diseases \cite{li2009two}.

% gioi thieu ve shared-task
The SOMD 2024 shared-task, hosted within Natural Scientific Language Processing and Research Knowledge Graphs (NSLP 2024) workshop \cite{NSLP2024}, is designed to extract mentioned software and metadata from documents. In this context, both the software and the metadata are identified as specific intervals in the original documents. Understand and identify the software mentioned in documents, which is especially important to support information extraction in scientific documents.

In this paper, we present three different approaches to address the challenge of sub-task I, including:

\begin{itemize}
    \item \textbf{Approach 1}: Fine-tuning pre-trained language models as a token classification problem.
    \item \textbf{Approach 2}: Two-stage framework for entity extraction and classification.
    \item \textbf{Apporach 3}: Three-stage framework for entity sentence classification, entity extraction, and entity type classification.
\end{itemize}

\section{Related Work}
In recent years, pre-training language models (PLMs) have made significant advancements in Named Entity Recognition (NER) tasks \cite{zhang2023samsung}. Among these, the most popular model is BERT \cite{devlin2018bert} and its variations like SciBERT \cite{beltagy2019scibert}, RoBERT \cite{chen2023robert}, and BiLSTM \cite{luo2018attention}. These models are often paired with machine learning techniques, particularly Conditional Random Fields (CRF) \cite{lopez2021mining}. Additionally, some approaches involve breaking down the NER task into two simpler tasks using question-answering methods \cite{arora2023split}, achieving notable results on various datasets like BioNLP13CG, CTIReports, OntoNotes5.0 \cite{pradhan2013ontonotesv5}, and WNUT17 \cite{WNUT2017} based on the F1 measure. 

With the emergence of ChatGPT, researchers have been exploring the use of Large Language Models (LLMs) for NER tasks \cite{wang2023gpt,zhang2024linkner}, with some studies demonstrating that ChatGPT can be distilled into smaller UniversalNER models for open NER \cite{zhou2024universalner}. These UniversalNER models have shown exceptional accuracy across 43 datasets spanning diverse fields such as biomedicine, programming, social media, law, and finance, without requiring direct supervision. UniversalNER surpasses traditional guideline-tuned models like Alpaca and Vicuna by an average of over 30 F1 points and achieves a high F1 score of 0.8 on SoMeSci. In this paper, BERT, SciBERT, and XML-R models are still utilized to address the first task of the shared SOMD 2024 challenge.

\section{Approach}
\begin{figure}[t]
\includegraphics[width=\textwidth]{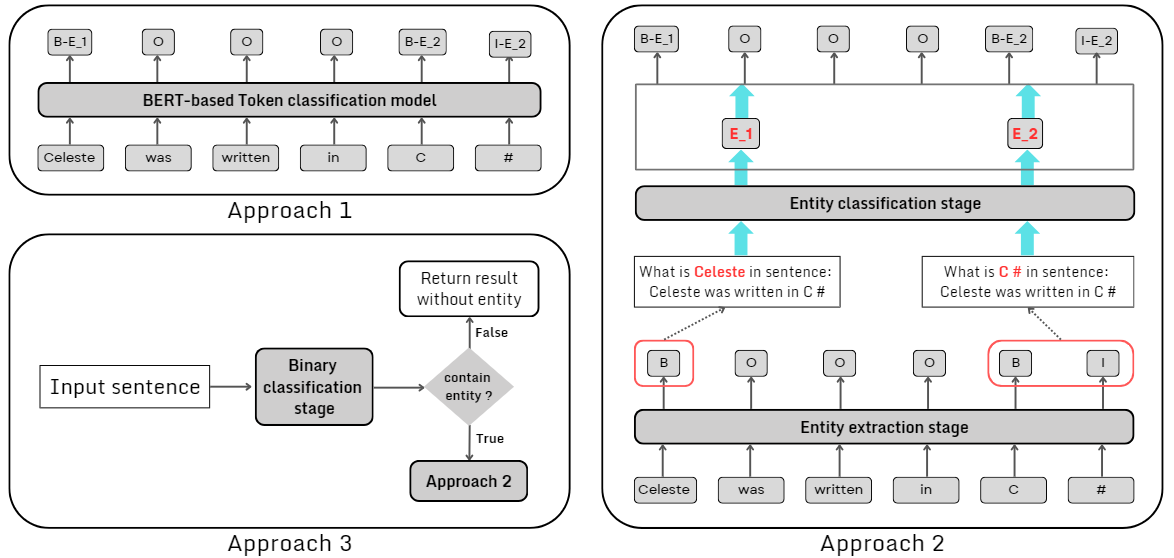}
\caption{\centering Overview system of three approaches: Sample input is "Celeste was written in C \#" with two entities are E\_1 and E\_2. E\_1 and E\_2 play the role of two entity types in this example } 
\label{fig2}
\end{figure}

To address the Software Mention Recognition task, we utilize the power of different pre-trained transformer-based language model in different approaches. Figure \ref{fig2} illustrates three approaches to participate in the competition. Because shared-task is related to each token in the sentence and whether words are in capital letters or not also greatly affects the recognition of entities. Therefore, we do not apply any preprocessing techniques but use data directly from the organizers. Also the tokenize method will depend on the default tokenier of the models. In our work, we employ various pre-trained language models, including the XLM-Roberta (XLM-R) \cite{Conneau2019UnsupervisedCR}, BERT \cite{devlin2018bert}, and SciBERT \cite{beltagy2019scibert} as our main backbones. The detail of our three approaches are present as follow.

\subsection{Approach 1: Token classification with BERTs}
For the first approach, we address the task by fine-tuning different transformer BERT-base models for the token classification task. We adapted different pre-trained language models to the training dataset. After tokenizing the input, we feed the token sequence to backbones models to extract the fixed vector in the last layer as the final representation of the input sentence. Then, we apply a fully connected layer to process the vectors and predict labels for each input token using a softmax function. There are a total of 27 labels (in Table \ref{labels_token}), where 26 correspond to 13 different entity types, and one label represents non-entities. Figure \ref{fig2} illustrates the overview of our first approach.
\begin{table}[t]
\centering
\caption{List of labels for token classification task in Approach 1}
\label{labels_token}
\resizebox{\textwidth}{!}{%
\begin{tabular}{|c|l|c|l|}
\hline
\textbf{Index} & \textbf{Label}                      & \textbf{Index}                 & \textbf{Label}                             \\ \hline
1     & B-Application\_ Creation   & 15                    & B-PlugIn\_Deposition              \\ \hline
2     & I-Application\_ Creation   & 16                    & I-PlugIn\_Deposition              \\ \hline
3     & B-Application\_Deposition  & 17                    & B-PlugIn\_Mention                 \\ \hline
4     & I-Application\_Deposition  & 18                    & I-PlugIn\_Mention                 \\ \hline
5     & B-Application\_Mention     & 19                    & B-PlugIn\_Usage                   \\ \hline
6     & I-Application\_Mention     & 20                    & I-PlugIn\_Usage                   \\ \hline
7     & B-Application\_Usage       & 21                    & B-ProgrammingEnvironment\_Mention \\ \hline
8     & I-Application\_Usage       & 22                    & I-ProgrammingEnvironment\_Mention \\ \hline
9     & B-OperatingSystem\_Mention & 23                    & B-ProgrammingEnvironment\_Usage   \\ \hline
10    & I-OperatingSystem\_Mention & 24                    & I-ProgrammingEnvironment\_Usage   \\ \hline
11    & B-OperatingSystem\_Usage   & 25                    & B-SoftwareCoreference\_Deposition \\ \hline
12    & I-OperatingSystem\_Usage   & 26                    & I-SoftwareCoreference\_Deposition \\ \hline
13    & B-PlugIn\_Creation         & 27                    & O                                 \\ \hline
14    & I-PlugIn\_Creation         & \multicolumn{1}{l|}{} &                                   \\ \hline
\end{tabular}
}
\end{table}

\subsection{Approach 2: Two-stage framework for Entity Extraction and Classification}

Motivated by recent work by \cite{arora2023split}, we address Task 1 - Software Mention Recognition with a two-stage framework composed of entity extraction and entity classification components. However, our components are re-designed to improve the overall performance than original framework proposed by \cite{arora2023split}. Figure 1 illustrates the overview of this approach, the detail of each component is presented below:

\begin{itemize}
    %\item \textbf{Stage 1- Entity detection}: The purpose of this stage is to detect tokens in a given input sentence whether they are entities or not. Essentially, we carry out this stage by token classification, similar to how we solve approach 1. The only difference is that instead of having a total of 27 labels of token like the above approach, we only use 3 labels: O, B, and I. Where label O corresponds to tokens that are not entities, B corresponds to tokens that are the first word of an entity (corresponding to 13 label types with the prefix B), and I corresponds to tokens inside an entity (corresponding to 13 label types with the prefix I). Using labels B and I to differentiate the position of tokens within an entity enables us to easily extract all words within an entity to proceed to stage 2.

    \item \textbf{Stage 1 - Entity extraction}: This stage aims to identify whether each token in a given input sentence belongs to an entity or not. We achieve this through token classification, similar to Approach 1. However, instead of using 27 labels for different token types, we only use 3 labels as:
    \begin{itemize}
        \item \textbf{O}: Non-entity token
        \item \textbf{B-X}: Beginning token of an entity of type X (where X represents one of the 13 entity types)
        \item \textbf{I-X}: Token within an entity of type X
    \end{itemize}
    
    Using separate labels for the beginning (B) and inside (I) positions of tokens within an entity allows us to efficiently extract all words belonging to the same entity in stage 2.

    \item \textbf{Stage 2 - Entity classification:} In this stage, we classify the detected entities from stage 1. We use a classifier with 13 labels corresponding to the 13 entity types, discarding the B-I prefix distinction used for token position. This classifier is built by fine-tuning a transformer-based model like BERT. 

    \cite{Tunstall_Werra_Wolf_2022} During fine-tuning for classification tasks, it's common practice to use the hidden state associated with the [CLS] token as input for a classifier. However, in this approach, we fine-tune the entire transformer model end-to-end. This means the hidden states are not treated as fixed features, but are trained alongside the classification head (a component added on top of the pre-trained model) for optimal performance. Additionally, to leverage the knowledge of transformer models, we format this classifier as a question-and-answering model by constructing the input as the following prompt:
    \begin{itemize}
        \item \textbf{Input}: What is $<$entity$>$ in the sentence: $<$input sentence$>$
        \item \textbf{Output}: Type of entity
    \end{itemize}

    %\item \textbf{Stage 2 - Entity classification:} With each entity detected in stage 1, we will feed it to a classifier to define its labels. This classifier will have 13 labels corresponding to 13 types of entities, instead of 26 labels, as we disregard the distinction between prefix B and I in each type of entity. We built this classifier by fine-tuning a transformer BERT-based model. When fine-tuning transformer models for classification tasks, it is popular to use the hidden stage associated with the [CLS] token as the input feature to build a classifier. In this study, we fine-tune a transformer model end-to-end for a classification task. It means that we do not use the hidden states as fixed features, but instead train them along with the classification head of the transformer model. The classification head is a component on top of the pre-trained model outputs, which can be easily trained with the base model. Additionally, to leverage the knowledge of transformer models, we format this classifier as a question-and-answering model by constructing the input as the following prompt:
    %\begin{itemize}
    %    \item Input: What is $<$entity$>$ in sentence: $<$input sentence$>$
    %    \item Output: Type of entity
    %\end{itemize}
    
\end{itemize}

\subsection{Approach 3: Three-stage framework}
Our analysis in Table \ref{general_statistics} revealed a limited number of sentences containing entities within the training set. This disparity raised concerns about potential biases in the label information during the training process for the previously mentioned approaches. To address this, we introduce a new three-stage framework, which integrate a binary classification with Approach 2.  We simply built a binary classification model to detect the sentences which contain the entity. As shown in Figure \ref{fig2}, if a sentence is classified as class 0, assign all tokens in the sentence as O, otherwise, this sentence will be passed to Approach 2 to extract the entity and its type.

\section{Experimental Setup}
\subsection{Data and Evaluation Metrics}
This shared-task uses the SoMeSci dataset \cite{SoMeSci} which included 39768 sentences and 3756 software mentions divided into a training set and a private test set. We train our systems only on the training set and evaluate the performance of our model on the private test set using weighted precision, recall, and F1-score.
In Table \ref{general_statistics}, we summarize some general information about the two data sets. Where \#Sentence denotes the number of sentences, \#Sentence with entity denotes the number of sentences containing the entity, and Total entity is the total of entities in all sentences. Max length and Avg length are the maximum length and average length of the sentences in each set, respectively.
This dataset contains six groups of entity Application, OperatingSystem, PlugIn, ProgrammingEnvironment, and SoftwareConference. Each group can have the entity belong to four types [Creation, Deposition, Mention, Usage]. In Table \ref{entity_statistics} we indicate the distribution of each entity in the dataset  
\begin{table}[t]
\centering
\caption{General statistics in the training set and private test set}
\label{general_statistics}
\begin{tabular}{|l|c|c|}
\hline
\textbf{Information}   & \textbf{Training set} & \textbf{Private test set} \\ \hline
\#Sentence             & 39768                 & 8180                      \\ \hline
\#Sentence with entity & 2353                  & 374                       \\ \hline
Total entity         & 3241                  & 515                       \\ \hline
Max length             & 568                   & 347                       \\ \hline
Avg length             & 28.32                 & 28.82                     \\ \hline
\end{tabular}
\end{table}

% Please add the following required packages to your document preamble:
% \usepackage{multirow}
\begin{table}[t]
\caption{\centering Statistics the number of  entities in each  entity type entity in each entity group in the training set and private test set}
\label{entity_statistics}
\resizebox{\textwidth}{!}{%
\begin{tabular}{|l|l|cc|cc|}
\hline
\multirow{2}{*}{\textbf{Entity group}}           & \multirow{2}{*}{\textbf{Entity}}         & \multicolumn{2}{c|}{\textbf{Training set}}                     & \multicolumn{2}{c|}{\textbf{Testing set}}                \\ \cline{3-6} 
                                        &                                 & \multicolumn{1}{c|}{\textbf{Quantity}} & \textbf{Total}                 & \multicolumn{1}{c|}{\textbf{Quantity}} & \textbf{Total}                \\ \hline
\multirow{4}{*}{Application}            & Application\_Creation           & \multicolumn{1}{c|}{150}      & \multirow{4}{*}{2353} & \multicolumn{1}{c|}{47}       & \multirow{4}{*}{348} \\ \cline{2-3} \cline{5-5}
                                        & Application\_Deposition         & \multicolumn{1}{c|}{80}       &                       & \multicolumn{1}{c|}{22}       &                      \\ \cline{2-3} \cline{5-5}
                                        & Application\_Mention            & \multicolumn{1}{c|}{162}      &                       & \multicolumn{1}{c|}{31}       &                      \\ \cline{2-3} \cline{5-5}
                                        & Application\_Usage              & \multicolumn{1}{c|}{1958}     &                       & \multicolumn{1}{c|}{248}      &                      \\ \hline
\multirow{2}{*}{OperatingSystem}        & OperatingSystem\_Mention        & \multicolumn{1}{c|}{13}       & \multirow{2}{*}{140}  & \multicolumn{1}{c|}{17}       & \multirow{2}{*}{33}  \\ \cline{2-3} \cline{5-5}
                                        & OperatingSystem\_Usage          & \multicolumn{1}{c|}{127}      &                       & \multicolumn{1}{c|}{16}       &                      \\ \hline
\multirow{4}{*}{PlugIn}                 & PlugIn\_Creation                & \multicolumn{1}{c|}{53}       & \multirow{4}{*}{344}  & \multicolumn{1}{c|}{17}       & \multirow{4}{*}{81}  \\ \cline{2-3} \cline{5-5}
                                        & PlugIn\_Deposition              & \multicolumn{1}{c|}{21}       &                       & \multicolumn{1}{c|}{8}        &                      \\ \cline{2-3} \cline{5-5}
                                        & PlugIn\_Mention                 & \multicolumn{1}{c|}{40}       &                       & \multicolumn{1}{c|}{11}       &                      \\ \cline{2-3} \cline{5-5}
                                        & PlugIn\_Usage                   & \multicolumn{1}{c|}{230}      &                       & \multicolumn{1}{c|}{45}       &                      \\ \hline
\multirow{2}{*}{ProgrammingEnvironment} & ProgrammingEnvironment\_Mention & \multicolumn{1}{c|}{41}       & \multirow{2}{*}{372}  & \multicolumn{1}{c|}{6}        & \multirow{2}{*}{49}  \\ \cline{2-3} \cline{5-5}
                                        & ProgrammingEnvironment\_Usage   & \multicolumn{1}{c|}{331}      &                       & \multicolumn{1}{c|}{43}       &                      \\ \hline
SoftwareCoreference                     & SoftwareCoreference\_Deposition & \multicolumn{1}{c|}{35}       & 35                    & \multicolumn{1}{c|}{4}        & 4                    \\ \hline
\end{tabular}
}
\end{table}

\subsection{System Settings}
We conduct all experiments on three approaches, using three base-version backbones: XLM-R\footnote{https://huggingface.co/FacebookAI/xlm-roberta-base}, BERT\footnote{https://huggingface.co/google-bert/bert-base-uncased}, and SciBERT\footnote{https://huggingface.co/allenai/scibert-scivocab-uncased}. We loaded the weights of the backbones from the HuggingFace library and carried out training on an NVIDIA T4(x2) GPU provided by Kaggle. The corresponding hyper-parameters for each approach are presented below:

\begin{itemize}
    \item \textbf{Approach 1: } batch size = 32, learning rate = 5e-05, and the number of epoch = 25 with XLM-R model and the number of epoch = 20 both remain backbones.
    \item \textbf{Approach 2: }
    \begin{itemize}
        \item Stage 1: batch size = 32, learning rate = 5e-05 and the number of epoch = 20 for all three backbones.
        \item Stage 2: batch size = 16, learning rate = 2e-05 and the number of epoch = 25 with XLM-R model and epoch = 20 two remainder models.
    \end{itemize}
    \item \textbf{Approach 3: }
    \begin{itemize}
        \item Stage 1: batch size = 32, learning rate = 2e-5 and the number of epoch = 10 for all three backbones.
        \item Stage 2 and Stage 3: Using the configuration and architecture as the Approach 2. 
    \end{itemize}
\end{itemize}

\section{Main results}
\label{section5}
According to the organizing committee, this sub-task will be evaluated by F1-Score based on exact matches. As shown in Table \ref{experiment}, we provide a tabulated summary of 9 experiments, each representing one of the 9 final systems generated from three different approaches and using three distinct backbones.

The experimental results in Table \ref{experiment} indicate that Approach 3, a three-stage system, demonstrates the best performance across all backbones, with the XLM-RoBERTa backbone exhibiting the highest efficacy among all approaches.  However, this result is for reference only and is only true in all of my experiments. It's important to acknowledge that different contexts, set up or datasets might yield different outcomes, and we are not sure this is the best result that each backbone could give in other cases. Finally, the best system was built according to approach 3 with XLM-R backbone and our best submission was ranked 3rd. Table \ref{scoreboard} show the final score of the top 5 participants.
\begin{table}[t]
\centering
\caption{\centering Comparative performance of our three Approaches with different pre-trained language models on the test set.}
\label{experiment}
\resizebox{\textwidth}{!}{%
\begin{tabular}{lccccccccc}
\hline
\multirow{2}{*}{\textbf{Models}} & \multicolumn{3}{c}{\textbf{Approach 1}} & \multicolumn{3}{c}{\textbf{Approach 2}} & \multicolumn{3}{c}{\textbf{Approach 3}} \\ \cline{2-10} 
 & \textbf{Precision} & \textbf{Recall} & \textbf{F1-score} & \textbf{Precision} & \textbf{Recall} & \textbf{F1-score} & \textbf{Precision} & \textbf{Recall} & \textbf{F1-score} \\ \hline
BERT & 0.675 & 0.594 & 0.625 & 0.682 & 0.643 & 0.653 & 0.690 & 0.629 & 0.650 \\ \hline
SciBERT & 0.658 & 0.621 & 0.623 & 0.719 & 0.645 & 0.670 & 0.736 & 0.631 & 0.670 \\ \hline
XLM-R & 0.716 & 0.614 & \textbf{0.649} & 0.707 & 0.654 & \textbf{0.671} & 0.729 & 0.649 & \textbf{0.678} \\ \hline
\end{tabular}%
}
\end{table}

\begin{table}[t]
\centering
\caption{Official scoreboard\protect\footnotemark for the sub-task I: Software mention recognition.}
\label{scoreboard}
\begin{tabular}{|l|c|ccc|}
\hline
\multirow{2}{*}{\textbf{Participant}} & \multirow{2}{*}{\textbf{Ranking}} & \multicolumn{3}{c|}{\textbf{Evaluation metrics}} \\ \cline{3-5} 
 &  & \textbf{Precision} & \textbf{Recall} & \textbf{F1-score} \\ \hline
phinx & Top 1 & \textbf{0.761} & \textbf{0.750}  & \textbf{0.740} \\
david-s477 & Top 2 &  0.739 & 0.711  & 0.692 \\
ottowg & Top 4 & 0.679 & 0.664 & 0.652 \\
vampire & Top 5 & 0.682 & 0.637 & 0.648 \\ \hline
Our best system & Top 3 & 0.729 & 0.649 & 0.678 \\ \hline
\end{tabular}
\end{table}
\footnotetext{https://codalab.lisn.upsaclay.fr/competitions/16935\#results}

\begin{comment}
With the test dataset labels provided by the organizing committee, we evaluated the performance of our best system for each entity class in Table \ref{entity_type}. From this Table, we observed that the SoftwareCoreference\_Deposition entity achieved the highest Precision score, while the ProgrammingEnvironment\_Usage entity attained the highest Recall and F1 score, top 5 F1-score classes are ProgrammingEnvironment\_Usage, SoftwareCoreference\_Deposition, and OperatingSystem\_Mention.It is evident that entities belonging to the PlugIn category typically scored lower than those in other categories, particularly PlugIn\_Deposition, where all metrics were 0.
\end{comment}

With the test dataset labels provided by the organizing committee, we evaluated the performance of our best system for each entity class in Table \ref{entity_type}. We observed that the SoftwareCoreference\_Deposition entity achieved the highest Precision score, while the ProgrammingEnvironment\_Usage entity attained the highest Recall and F1 score, top 5 F1-score classes are ProgrammingEnvironment\_Usage, SoftwareCoreference\_Deposition, and OperatingSystem\_Mention.
It is evident that entities belonging to the PlugIn group typically scored lower than those in other groups shows that it has difficulty in the regconization process. Although, the number of PlugIn\_Usage entities in the training set is pretty large the result on the test set is not positive. Besides that, PlugIn\_Creation and PlugIn\_Deposition entities have the sample in the training set are pretty low and their score moves forward to zero. The number of OperatingSystem\_Mention entities in the training set is low and the score on the test set is high so we predict the mention entity type in this group is featured and easier to recognize than other groups.

Additionally, in Table \ref{stage}, we evaluated each individual stage in our final three-stage system by assuming that the accuracy of the stages before it is 100\%. The first stage works well with an F1-score of 0.992 in classifying whether a sentence contains an entity or not. Moving to stage 2, tasked with detecting entities in sentences, achieved an F1 score at a relatively good level, but a significant difference between Precision and Recall (12.6\% difference) is evident, which also affects the overall system performance. In the final stage, the scores between the three metrics are relatively balanced, but it appears that the task of classifying 13 entity classes had some impact on this stage with relatively lower overall performance. The propagation of errors between the three stages has a significant impact on the entire system, with the final F1-score of the entire system being 0.678.

\begin{table}[t]
\centering
\caption{\centering Performance of the final system on the test dataset across entity classes evaluated by Precision, Recall, and F1-score.}
\label{entity_type}
\begin{tabular}{|l|c|c|c|}
%\toprule
\hline
\textbf{Entity class} & \textbf{Precision} & \textbf{Recall} & \textbf{F1-score} \\ %\midrule
\hline
Application\_Creation            & 0.692   & 0.766  & 0.727   \\
Application\_Deposition          & 0.615   & 0.727  & 0.667   \\
Application\_Mention             & 0.560   & 0.452  & 0.500   \\
Application\_Usage               & 0.812   & 0.730  & 0.769   \\
OperatingSystem\_Mention         & 0.867   & 0.765  & 0.812   \\
OperatingSystem\_Usage           & 0.579   & 0.688  & 0.629   \\
PlugIn\_Creation                 & 0.200   & 0.059  & 0.091   \\
PlugIn\_Deposition               & 0.000   & 0.000  & 0.000   \\
PlugIn\_Mention                  & 0.667   & 0.364  & 0.471   \\
PlugIn\_Usage                    & 0.682   & 0.333  & 0.448   \\
ProgrammingEnvironment\_Mention  & 0.500   & 0.167  & 0.250   \\
ProgrammingEnvironment\_Usage    & 0.886   & \textbf{0.907}  & \textbf{0.897}   \\
SoftwareCoreference\_Deposition  & \textbf{1.000}   & 0.750  & 0.857   \\ %\bottomrule
\hline
\end{tabular}
\end{table}

\begin{table}[t]
\centering
\caption{\centering Performance of components in our final three-stage framework.}
\label{stage}
\begin{tabular}{|c|ccc|}
\hline
\textbf{Stage} & \textbf{Precision} & \textbf{Recall} & \textbf{F1-score} \\ \hline
Stage 1         & 0.992               & 0.992            & 0.992             \\
Stage 2         & 0.912               & 0.786            & 0.845             \\
Stage 3         & 0.786               & 0.806            & 0.784             \\ \hline
\end{tabular}
\end{table}

\section{Conclusion and Future Work}
In this paper, we present and evaluate three approaches for tackling sub-task I in the Software Mention Detection in Scholarly Publications shared task. While we explored the use of suitable transformer models like BERT, our three-stage system leveraging the XLM-R model achieved the highest performance in the competition. As a result, our best system achieved the Top 3 in the private test. In future work, our intention is to analyze the error propagation between the three stages to enhance the performance of the entire three-stage system. Additionally, with access to more substantial computational resources, we aim to experiment with fine-tuning sub-tasks using larger batch sizes and epochs for each backbone in order to investigate the effects of these hyper-parameters on the model's performance.

\section*{Acknowledgements}
This research was supported by The VNUHCM-University of Information Technology's Scientific Research Support Fund. We also thank the anonymous reviewers for their valuable comments on our manuscript. 
%
% ---- Bibliography ----
%
% BibTeX users should specify bibliography style 'splncs04'.
% References will then be sorted and formatted in the correct style.
%
\bibliographystyle{splncs04}
\bibliography{mybibliography}
\end{document}